# Enhancing Flood Forecasting with Dual State-Parameter Estimation and Ensemble-based SAR Data Assimilation

Thanh Huy Nguyen[1], Sophie Ricci[1], Andrea Piacentini[1], Raquel Rodriguez Suquet[2], Gwendoline Blanchet[2], Santiago Peña Luque[2] and Peter Kettig[2]

[1]: CECI, CERFACS/CNRS UMR 5318, Toulouse, 31057, France
[2]: Centre National d'Etudes Spatiales (CNES), Toulouse, 31401, France

thnguyen@cerfacs.fr

*Abstract* – **Ensemble data assimilation in flood forecasting depends strongly on the density, frequency and statistics of errors associated with the observation network. This work focuses on the assimilation of 2D flood extent data, expressed in terms of wet surface ratio, in addition to the in-situ water level data. The objective is to improve the representation of the flood plain dynamics with a TELEMAC-2D model and an Ensemble Kalman Filter (EnKF). The EnKF control vector is composed of friction coefficients and corrective parameters to the input forcing. It is augmented with the water level state averaged over selected subdomains of the floodplain. This work focuses on the 2019 flood event that occurred over the Garonne Marmandaise catchment. The merits of assimilating SAR-derived flood plain data complementary to in-situ water level observations are shown in the control parameter and observation spaces with 1D and 2D assessment metrics. It was also shown that the assimilation of Wet surface Ratio in the flood plain complementary to in-situ data in the river bed brings significative improvement when a corrective term on flood plain hydraulic state is included in the control vector. Yet, it has barely no impact in the river bed that is sufficiently well described by in-situ data. We highlighted that the correction of the hydraulic state in the flood plain significantly improved the flood dynamics, especially during the recession. This proof-of-concept study paves the way towards near-real-time flood forecast, making the most of remote sensing-derived flood observations.**

## I. INTRODUCTION

The occurrence and intensity of natural disasters—among which flooding is one of the most common and costliest—has increased over recent decades, especially in the context of climate changes. In 2021 alone, the Emergency Event Database (EM-DAT) recorded 432 disastrous events related to natural hazards worldwide with 223 flood events having affected more than 100 million people and accounted for an economic loss of 74 billion USD. While hydrology and hydraulic numerical models play an essential role in flood forecasting, their capabilities remain limited due to uncertainties in their input data such rainfall, inflow, geometry of the catchment and the river (e.g. topographic errors from Digital Elevation Models (DEMs) and bathymetric errors) and hydraulic parameters (calibration of friction coefficients). Data assimilation is an efficient tool to reduce these uncertainties, by combining numerical model outputs with various Earth Observations from space or from in-situ measurements. The increasing volume of data from space missions provide heterogeneous and relevant data, such as altimetry (TOPEX/POSEIDON, Jason-1/2/3, ENVISAT, SARAL/ ALTIKA, Sentinel-3/6, SWOT), optical (Pléïades, Sentinel-2) and radar (Sentinel-1, TerraSAR-X). A conventional DA approach stands in the assimilation of water surface elevation data, either from in-situ time-series, from altimetry or retrieved from Synthetic Aperture Radar (SAR) images using river width information with complementary DEM data. An updated review from [1] provides the state-of-the-art on the assimilation of Earth Observation data with hydraulic models for the purpose of improved flood inundation forecasting.

SAR data provides an all-weather global imagery of continental waters depicted by low backscatter values resulting from the specular reflection of the incident radar pulses upon arrival at the water surfaces. While the assimilation of SAR-derived water level (WL) information is convenient as it deals with a diagnostic variable of the model, it depends on the usage of a DEM and thus may suffer from the lack of precision of high-resolution topographic data [2-4]. This constraint can be relaxed with the direct assimilation of SAR-derived flood probability maps or flood extent maps. Hostache *et al.* [5] presents the assimilation of SAR-derived flood probability maps with a Particle Filter (PF). For that matter, a probabilistic flood map is derived from SAR backscatter images using a Bayesian approach to assign a probability of flooded to its pixels, assuming that the prior probabilities for a backscatter value to be flooded or non-flooded follow two gaussian distributions, as detailed in [6]. Cooper *et al.* [7] proposed an observation operator that directly uses backscatter values from SAR images as observations in order to bypass the flood edge identification or flood probability estimation processes. Similarly to [5, 6], this approach also relies on the hypothesis that SAR images must yield distinct distributions of wet and dry backscatter values, which may not hold for real SAR data.

In this work, we propose to take into account the flood extent information derived from SAR S1 images, as wet surface ratios (WSR). This is the ratio between the number of wet pixels within a floodplain subdomain and the total number of pixels. An ensemble data assimilation (DA) approach to accommodate 2D WSR observations alongside in-situ water level time-series within an EnKF framework has been implemented on the TELEMAC-2D hydrodynamic model set up over the Garonne Marmandaise catchment. A dual state-parameter DA strategy is also carried out to reduce the uncertainties in friction coefficients, upstream forcing and hydraulic state considered as water level averaged over selected subdomains of the floodplain.

## II. Study Area and Model

### A. *Shallow Water Equations in TELEMAC-2D*

Free-surface hydraulic modelling is principally governed by the Shallow Water equations (SWE, also known as Saint-Venant equations derived from Navier-Stokes Equations), which express mass and momentum conservation averaged in the vertical dimension. In this work, the hydrodynamic numerical model TELEMAC-2D is used to simulate and predict the water level (denoted by $H$ [m]) and velocity (with horizontal components denoted by $u$ and $v$ [m.s$^{-1}$]) from which flood risks can be assessed. It solves the SWE with an explicit first-order time integration scheme, a finite-element scheme and an iterative conjugate gradient method. A complete description of the underlying theoretical approach is provided in [8].

The 2D domain is described by a triangular mesh, in which each node associates with a topographical height. Different parameters are defined, including the friction coefficients, using Strickler formulation [9] denoted by $K_s$, defined uniformly over of a number of segments of the riverbed and over the whole floodplain. The mesh is constructed with three distinguished mesh types: (i) the riverbed with an oriented and fine mesh (max. triangle size 80 m) which guides the flows; (ii) the floodplain with an unstructured and coarse mesh (max. triangle size 150 m); and (iii) the dyke systems modelled by guidelines along which the mesh is very fine (max. triangle size 40 m). Beside topographic and bathymetric data, hydraulic models require a time-varying hydrograph of the inflow discharge at the upstream boundary, initial conditions, and outflow WL data or a rating curve at the downstream boundary.

### B. *Study Area and Event*

The study area is the Garonne Marmandaise catchment (southwest France) which extends over a 50-km reach of the Garonne River between Tonneins and La Réole (Figure 1). This catchment has been equipped with infrastructures, and a progressively constructed system of dykes and weirs to protect the floodplains from flooding events such as the historic flood of 1875 and manage submersion and flood retention areas. Observing stations operated by the VigiCrue network (https://www.vigicrues.gouv.fr/) are located at Tonneins, Marmande, and La Réole (indicated as black circles).

A TELEMAC-2D model was developed and calibrated by EDF R&D [10] over this catchment, built on a mesh of 41,000 nodes using bathymetric cross-sectional profiles and topographic data [10]. A local rating curve at Tonneins is used to translate the observed WLs into a discharge time-series that is applied over the whole upstream interface (cyan arrow), including both river bed and floodplain boundary cells. Such a modeling strategy was implemented to allow for a cold start of the model with any inflow value. However, it prompts an artificial over-flooding of the upstream first meander, which remains for a period of time until the water returns to the river bed. On the other hand, the downstream BC at La Réole is described with a local rating curve built from gauge measurements. Over the simulation domain, the friction coefficient is defined over seven zones, $K_{s_1}$ to $K_{s_6}$ for the river bed and $K_{s_0}$ for the entire floodplain, as illustrated in Figure 1 with solid-colored segments of the river bed and white region for the floodplain [11, 12]. The description of the friction coefficients is highly prone to uncertainties related to the zoning assumption, the calibration procedure, and the set of calibration events. In the absence of in-situ data in every river segment, their a priori values are set based on the calibration process from the original design by EDF.

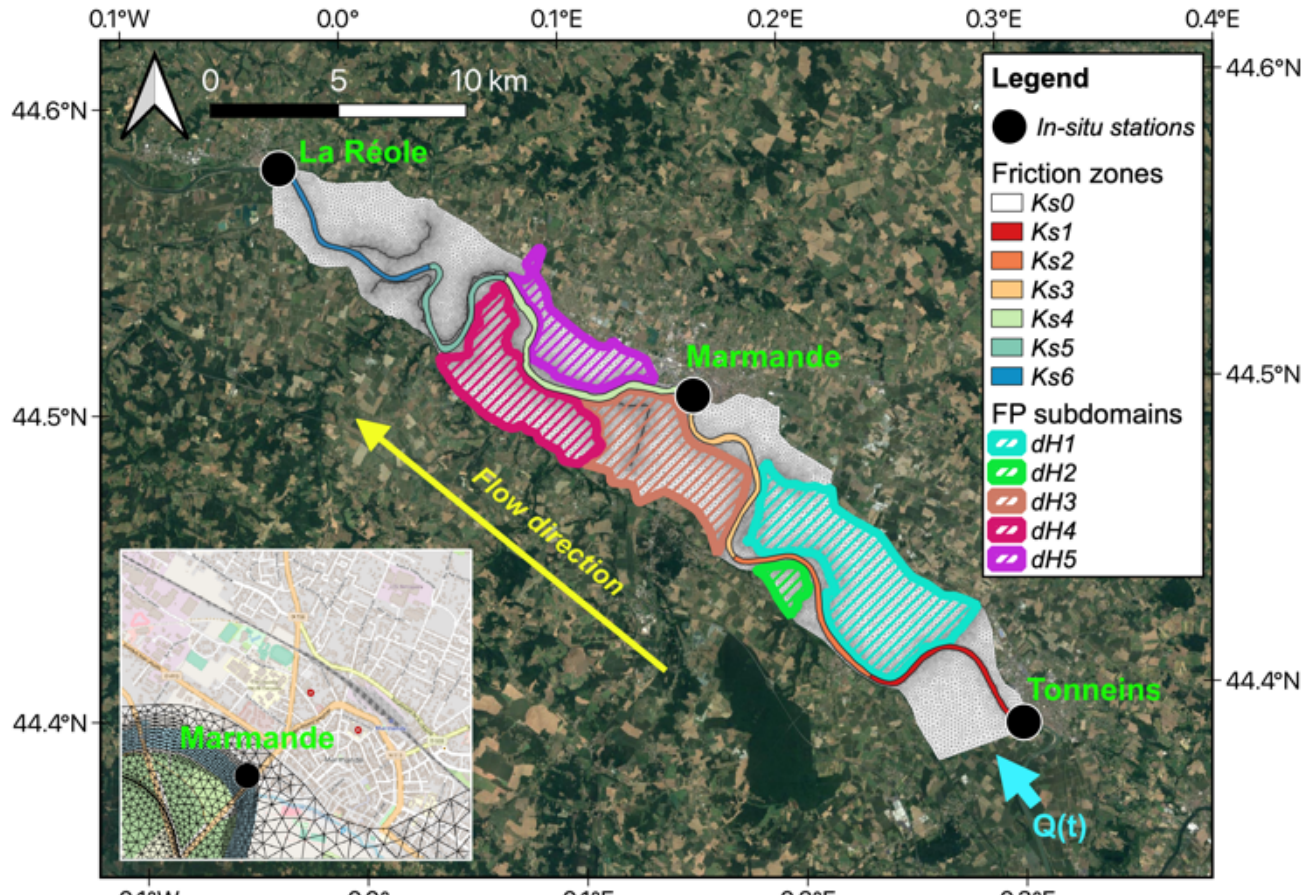

Figure 1. TELEMAC-2D Garonne Marmandaise model and control vector. Inset figure magnifies the impacted urban area around Marmande.

A significant flood event occurred in December 2019 has been studied in this research work. In-situ WL measured every 15 minutes at Tonneins, Marmande and La Réole from VigiCrue observing network are shown in Figure 2. This double-peak flood event was observed by eleven Sentinel-1 (S1) SAR images, provided by the constellation of S1-A and S1-B ascending and descending orbits, represented by the black vertical dashed lines. There are also two Sentinel-2 (S2) optical images available, represented by the red vertical dashed lines, near the first flood peak at 2019-12-15 12:05 and 2019-12-17 11:54 thus providing independent data for validation, with a cloud cover ratio of 40.6% and 11.3%, respectively. In this work, the S1-derived flood extent maps are used for the assimilation in combination with the in-situ WL observations whereas the S2-derived ones are only used for validation.

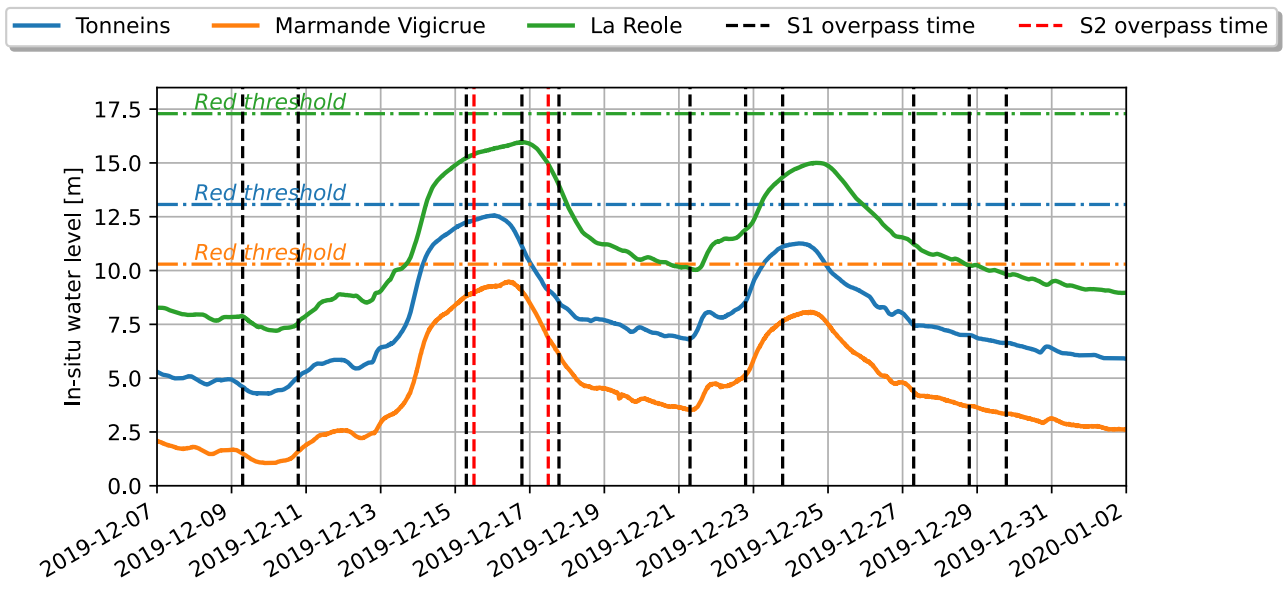

Figure 2. Water level time series at VigiCrue observing stations, and S-1 and S-2 overpass times.

## III. Method

### A. Data Assimilation

In this work, the WL time-series at the three VigiCrue stations (Tonneins, Marmande and La Réole) and the WSR computed over the five floodplain zones, are assimilated with the EnKF algorithm implemented on the TELEMAC-2D Garonne model. This allows for a sequential correction of the friction, the inflow discharge, and water level averaged over selected subdomains of the floodplain. For full description of the performed cycled EnKF, please refer to [13, 14].

### B. Control Vector

Table I summarizes the variables included in the control vector of the EnKF. The friction coefficients are considered as random variables with a gaussian Probability Density Function (PDF) with mean and standard deviation estimated from the calibration process. The uncertainty in the upstream BC is also taken into account. Indeed, the limited number of in-situ observations also yields errors in the formulation of the rating curve that is used to translate the observed WLs into discharge, especially for high flows. Therefore, a multiplicative factor $\mu$ applied on the time-dependent discharge time-series is considered as a random variable with a gaussian PDF centered at 1.

In addition, in order to account for the evapotranspiration, ground infiltration and rainfall processes that are lacking in the TELEMAC-2D Garonne model, a state correction is implemented in the floodplain. Five subdomains (hereafter called *zones*) delineated in the floodplain beyond the dykes, involving a uniform WL correction over each zone, are added to the control vector. These state corrections $\delta H_{[1:5]}$ are considered as zero-mean gaussian random vectors. These zones were determined based on the description of the storage areas [10] and the dyke system of the catchment. It is worth-noting that the first storage area of the model, at the first meander near Tonneins, is excluded in this study because of the aforementioned artificial over-flooding effect. In addition, several storage areas near the downstream area are not considered, because these areas are not fully observed by S1, and spurious dynamics may be caused by the errors in topography near La Réole [10]. Over each of the five zones, the WSR between the area of observed wet surfaces and the total area of the zone is computed.

Table I Gaussian PDF of uncertain input variables.

| Var-iables | Unit | Calibrated/ default value $x_0$ | Standard deviation $\sigma_x$ | 95% confidence interval |
|---|---|---|---|---|
| $K_{s_0}$ | $m^{1/3}s^{-1}$ | 17 | 0.85 | 17 ± 1.67 |
| $K_{s_1}$ | $m^{1/3}s^{-1}$ | 45 | 2.25 | 45 ± 4.41 |
| $K_{s_2}$ | $m^{1/3}s^{-1}$ | 38 | 1.9 | 38 ± 3.72 |
| $K_{s_3}$ | $m^{1/3}s^{-1}$ | 38 | 1.9 | 38 ± 3.72 |
| $K_{s_4}$ | $m^{1/3}s^{-1}$ | 40 | 2.0 | 40 ± 3.92 |
| $K_{s_5}$ | $m^{1/3}s^{-1}$ | 40 | 2.0 | 40 ± 3.92 |
| $K_{s_6}$ | $m^{1/3}s^{-1}$ | 40 | 2.0 | 40 ± 3.92 |
| $\mu$ | - | 1 | 0.06 | 1 ± 0.12 |
| $\delta H_{[1:5]}$ | $m$ | 0 | 0.25 | 0 ± 0.49 |

### C. Experimental Setup

One free run (FR) and three DA experiments (IDA, IWDA, IHDA) were carried out (Table 1) with different configurations regarding the types of observations that are assimilated and the active components of the control vector. Two types of observations are considered: (i) in-situ WL observations at three VigiCrue stations Tonneins, Marmande and La Réole every 15 minutes, (ii) WSR measurements on the five floodplain zones (corresponding to $\delta H_{[1:5]}$) at the eleven S1 overpass times (Figure 2).

Table II Summary of the Free Run and DA experiment settings.

| Exp. | DA | Assimilated observations | Nb of members | Control variables |
|---|---|---|---|---|
| FR | No | - | 1 | - |
| IDA | Yes | In-situ | 75 | $K_{s_0}, K_{s_{[1:6]}}, \mu$ |
| IWDA | Yes | In-situ and WSR | 75 | $K_{s_0}, K_{s_{[1:6]}}, \mu$ |
| IHDA | Yes | In-situ and WSR | 75 | $K_{s_0}, K_{s_{[1:6]}}, \mu, \delta H_{[1:5]}$ |

## IV. Results

In this section, quantitative performance assessments are carried out in the control and in the observation spaces by

(i) comparing the parameters yielded by the different DA analysis;

(ii) comparing the different analyzed WL time-series with synthetical or real in-situ observations;

(iii) comparing the different analyzed WSR with real or synthetical WSR observations in the floodplain;

(iv) evaluating the contingency maps and the overall Critical Success Index (CSI) and Cohen's kappa index ($\kappa$) with respect to the observed flood extent maps (S1-derived ones that were used to yield WSR, or S2-derived ones only used for validation). While CSI is conventionally the most widely used metric for this comparison, Cohen's kappa index provides a better overall metric with correctly predicted non-flooded pixels also being taken into account.

## A. Results in the control space

The analyzed parameters from the different DA experiments are shown in Figure 3 where horizontal dashed lines stand for the default values $\mathbf{x}_0$ (Table I), blue curves for IDA, green curves for IWDA, and red curves for IHDA. Vertical lines show the acquisition time of the S1 images, providing WSR observations being assimilated in the IWDA and IHDA experiments.

The analyzed values for $K_{s_k}$ (with $k \in [0,6]$) over the flood event are shown on the top panel in Figure 3. The analysis for the inflow correction $\mu$ and for $\delta H_k^a$ with $k \in [1,5]$ (only by IHDA) are shown on panel in Figure 3. The bottom panel of Figure 3b displays the upstream forcing for referential purposes. First, it should be noted that for all DA experiments, the analysis values for the friction coefficients in the river bed and the floodplain remain within physical ranges. The analysis for IHDA is closer to that of IWDA, compared to IDA as they both assimilate in-situ and WSR data. The analyses in the 4th friction segment (i.e. $K_{s_4}$), which includes the Marmande in-situ station, are relatively close for IDA, IWDA and IHDA, showing that the assimilation of in-situ WLs at Marmande suffices to account for friction errors in this area. For the friction coefficients of the 5th and 6th river segments (i.e. $K_{s_5}$ and $K_{s_6}$), the analysis is quite far from the calibrated values which is most likely due to the poor quality of the model topography in the downstream part of the domain, as well as the large misfit between the in-situ and the simulated WLs at La Réole. The analyses on $\mu$ are very similar among IDA, IWDA, and IHDA. This suggests that the in-situ WLs observed at upstream station Tonneins are enough to constrain the multiplicative correction to the inflow and that the use of additional data in the floodplain is unnecessary. Concerning the $\delta$H parameters controlled by IHDA, the mostly negative correction on all $\delta$H values increase (i.e. more water is removed in the corresponding floodplain zones) as the flood rises, especially at the flood peak, between the two peaks, and during the recess period in order to account for the TELEMAC-2D model's limitation in physical process to empty the floodplain.

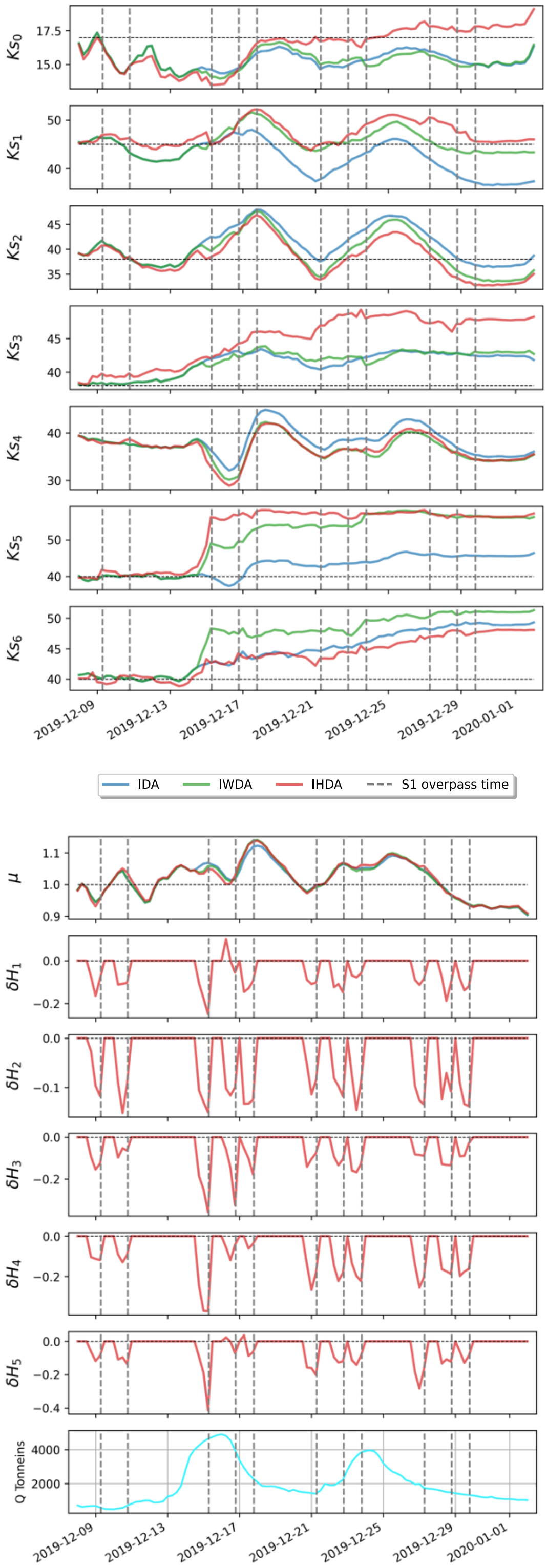

Figure 3. Evolution of controlled parameters for friction, multiplicative correction to the inflow, state correction in the floodplain. The last panel depicts the forcing inflow at Tonneins.

### *B. Results in the observation space: Water levels at observing stations*

Table III Water level RMSE w.r.t. in-situ WL at VigiCrue observing stations. Lowest RMSE values are underlined.

| **Exp.** | **RMSE [m]** | | |
|---|---|---|---|
| | ***Tonneins*** | ***Marmande*** | ***La Réole*** |
| FR | 0.129 | 0.220 | 0.318 |
| IDA | <u>0.060</u> | <u>0.045</u> | <u>0.125</u> |
| IWDA | 0.064 | 0.049 | 0.128 |
| IHDA | 0.064 | 0.051 | 0.138 |

The RMSEs computed over time for the 2019 event for the WLs from the FR, IDA, IWDA and IHDA, with respect to the observed WLs at Tonneins, Marmande, and La Réole are summarized in Table III. For each observing station, the lowest RMSE values are underlined, which show the slight advantage of IDA which only concerns the assimilation of the WLs at these observing stations. Figure 4 depicts the WLs simulated by FR (Figure 4a) and IHDA (Figure 4b) compared to the in-situ observations, in dashed curves. It is shown that all DA experiments succeed in significantly reducing the WL errors compared to those of FR. An important message is that the addition of WSR data does not bring significative improvement (nor does it degrade it) to the dynamics in the river bed that is already well described by in-situ data assimilation. The three DA experiment bring a significative improvement with respect to the Free run in the river bed. The reductions in RMSE with respect to FR amount to 50%, 77%, and 57%, respectively, at Tonneins, Marmande, and La Réole, with close values between IDA, IWDA, and IHDA. The RMSEs at Tonneins and Marmande remain under 6.5 cm for all DA experiments, whereas it is under 14 cm at La Réole.

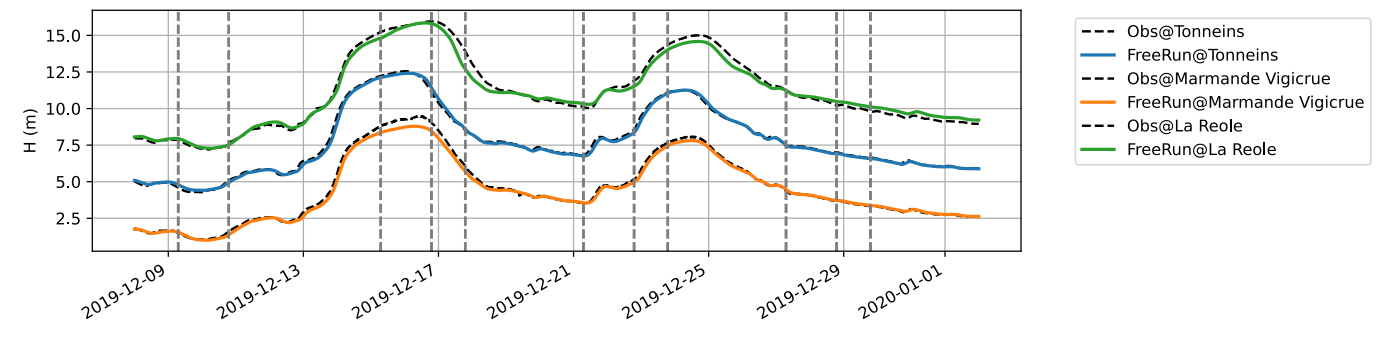

(a) FR (RMSE: 0.318, 0.129, 0.220)

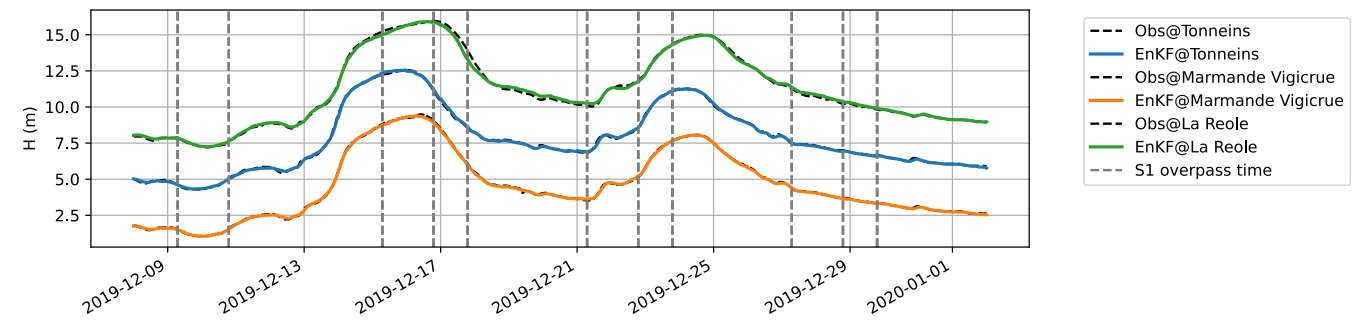

(b) IHDA (RMSE: 0.138, 0.064, 0.051)

Figure 4. Simulated WL compared to observations.

It should be noted that the best DA strategy according to in-situ WL RMSE is IDA (although slightly). The additional WSR observations in the floodplain (assimilated in IWDA and IHDA) leads to a smaller WL improvement from FR at observing stations than IDA does. It is highly probable that an extended control vector is necessary to account for model error in the river bed and in the flood plain. For instance, we could consider a finer zoning of friction in the floodplain, the addition of lateral tributaries that mainly carry a large volume of water for high flows, a more precise description of the topography in the floodplain, or an addition of physical processes in the SWE solver such as rain and evapotranspiration. A preliminary conclusion here is that the assimilation of data in the floodplain does not bring any significant improvement to the flood dynamics when only assessed in the river bed, through 1D metrics defined in the river bed.

### *C. Results in the observation space: WSR in the floodplain*

The performance of the DA strategy is now assessed in the observation space of WSR, in the floodplain. The WSR in the five floodplain subdomains for the simulated WL in FR and the analyzed WL in the three DA experiments with a wet threshold of 5 cm are compared to the WSR computed from the observed S1-derived flood extent maps. The misfit between simulation and observation WSR values (i.e. observed WSR - simulated WSR) are shown in Figure 5, which allows to assess the performance of the simulation in terms of flood extent representation. The color codes for the experiments remain the same as in previous figures, i.e., FR in orange, IDA in blue, IWDA in green, and IHDA in red.

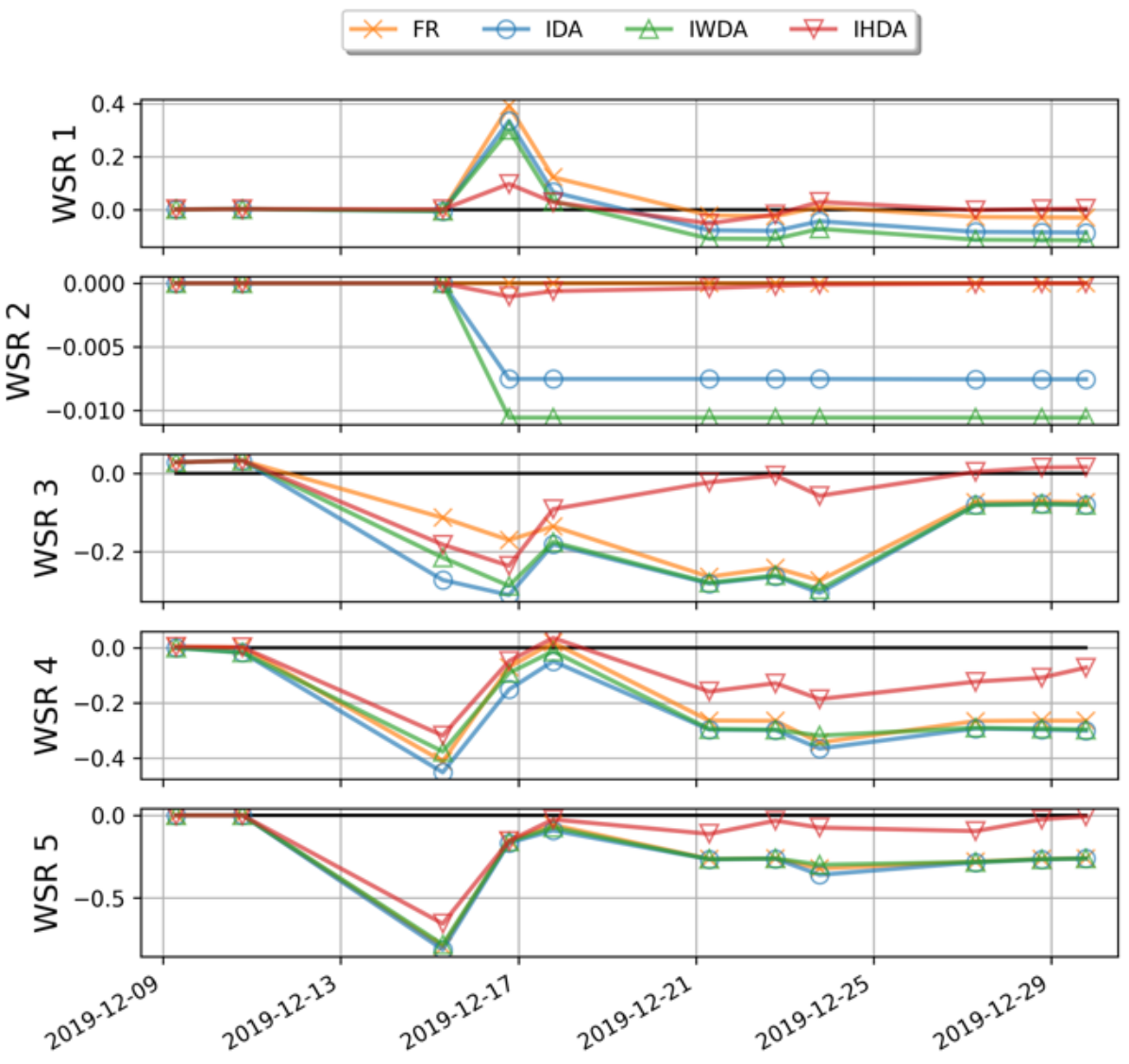

Figure 5. Errors between truth's WSR values and simulated WSR values in the 5 floodplain zones.

First, it can be noted that the analyses for IDA and IWDA do not bring much improvement in terms of flood extent representation with respect to FR. The improvement is much more evident for IHDA, as we can see it yields the smallest WSR misfits among the experiments. In particular, IHDA brings a significant improvement for the subdomains 1, 3, 4 and 5; whereas the misfits in subdomain 2 have already been small for FR, hence the contributions from IHDA are less obvious. Compared to IWDA, the assimilation of WSR by IHDA with the extended control vector brings an improvement in all subdomains, and thus allows the floodplain to be efficiently emptied after the flood peak.

A significant overprediction at the 3rd timestep, right before the first peak (2019-12-15 07:00), in subdomain 4 and 5 can be observed. This could stem from the characteristics of SAR backscatter which intensifies as the soil moisture increases due to rainfalls while the area has not been flooded.

At later moments during the flood, the correction of the hydraulic state in the floodplain for IHDA, during the recess of the first peak (between 2019-12-17 and 2019-12-21), allows for a better simulation of the second flood peak than in FR.

### *D. 2D validation with contingency maps, CSI and κ indices*

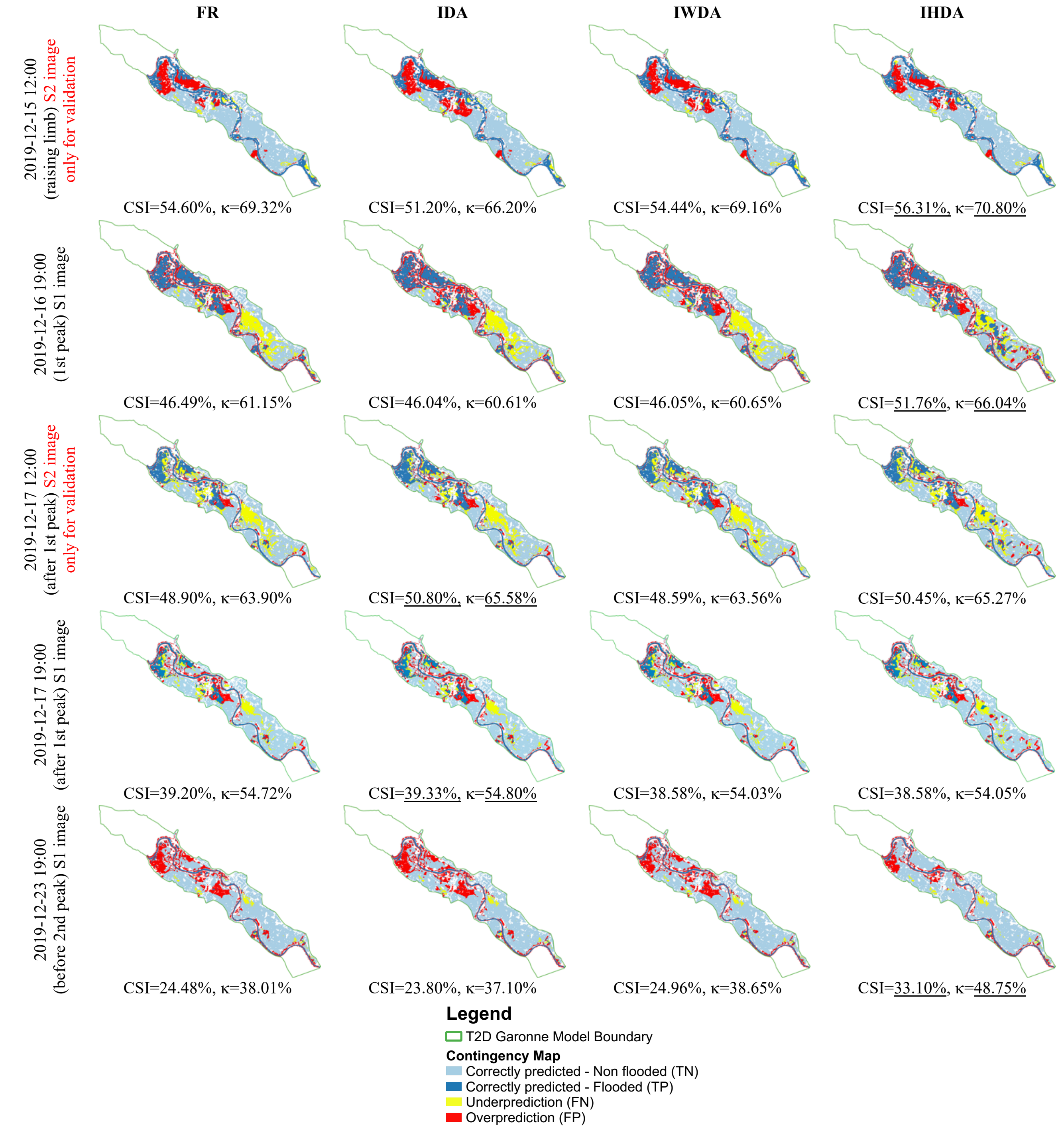

Figure 6. Contingency maps computed between simulated flood extent (from left to right: FR, IDA, IWDA and IHDA) with respect to S1-derived flood extent (row 2, 4 and 5) and S2-derived flood extent (rows 1 and 3). Highest CSI and κ scores are underlined.

2D validations are carried out by evaluating contingency maps comparing TELEMAC-2D water masks with S1-/S2-derived flood maps at their respective overpass times, and by quantitatively assessing the resulting CSI and the κ index scores. Figure 6 depicts the contingency maps based on the comparison of the TELEMAC-2D simulated flood extent maps from FR and DA experiments with respect to those derived from S1 or S2 images during the 2019 flood event. It should be stressed that S2 imagery data are not assimilated, they are only used for validation as independent data. The associated CSI and the κ indices are indicated on each contingency map. The contingency maps are shown from top to bottom, at satellite overpass time:

- at the rising limb of the first flood peak observed by S2 (2019-12-15 12:00),
- at first flood peak by observed S1 (2019-12-16 19:00),
- during the first falling limb observed by S2 (2019-12-17 12:00) and then by S1 (2019-12-17 19:00),
- before the second flood peak observed by S1 (2019-12-23 19:00).

From the first row in Figure 6, IHDA brings some improvements with respect to FR, IDA and IWDA before the flood peak, with relatively significant overprediction regions (red pixels) on subdomain 4 and 5 from all experiments can be observed on these first-row figures. It is coherent with the remark made on the WSR validation. At the first flood peak observed by S1 image (second row in Figure 6), IHDA allows better predictions of the flooded pixels (represented by dark blue pixels), mostly in subdomain 1. During the first flood recess (third and fourth row in Figure 6), the improvement brought by IHDA is not as visible as at the flood peak (second row). Relatively large amount of underprediction (yellow pixels) in the subdomains 1 and 3 remains significant which suggests a further improvement to be made concerning the topography and friction of these subdomains.

The added validation of the S2 image at 2019-12-17 12:00 (third row) provides an interesting remark. Indeed, the observed flood extent detected on this image is more similar to the one captured by the S1 image at 2019-12-16 19:00 (17 hours before) than to the one right afterward at 2019-12-17 19:00 (5 hours later). Such a non-linear situation, taking into account the fact that these three images were acquired in the span of 24 hours during the start of the falling limb, shows the different tendencies between the in-situ WL and the floodplain dynamics.

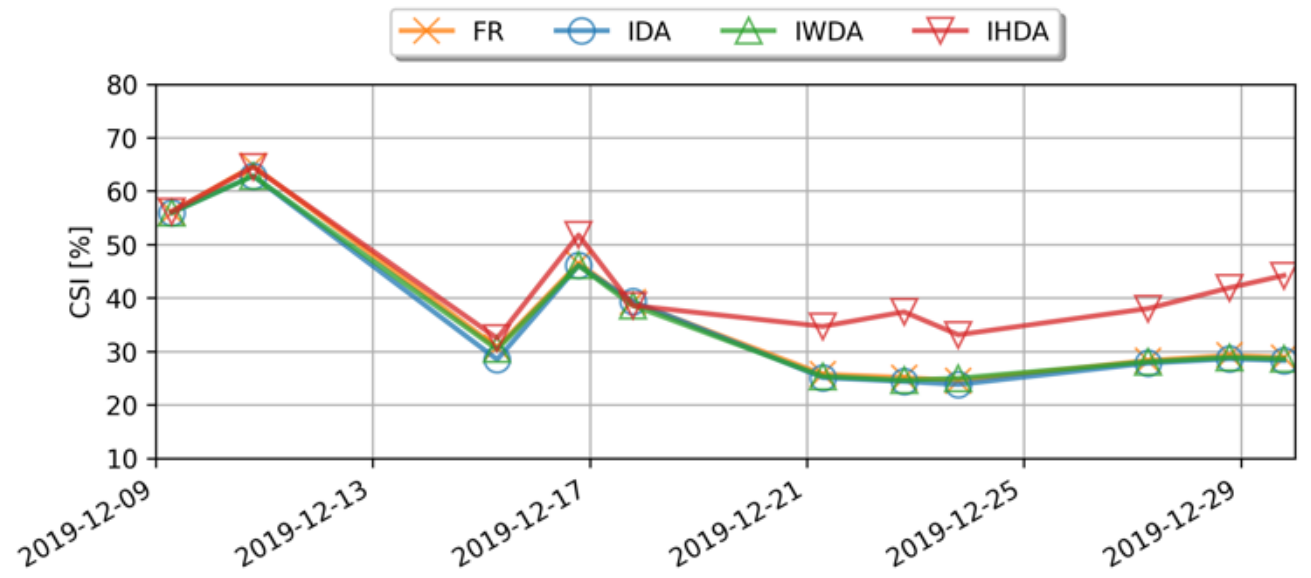

(a) CSI

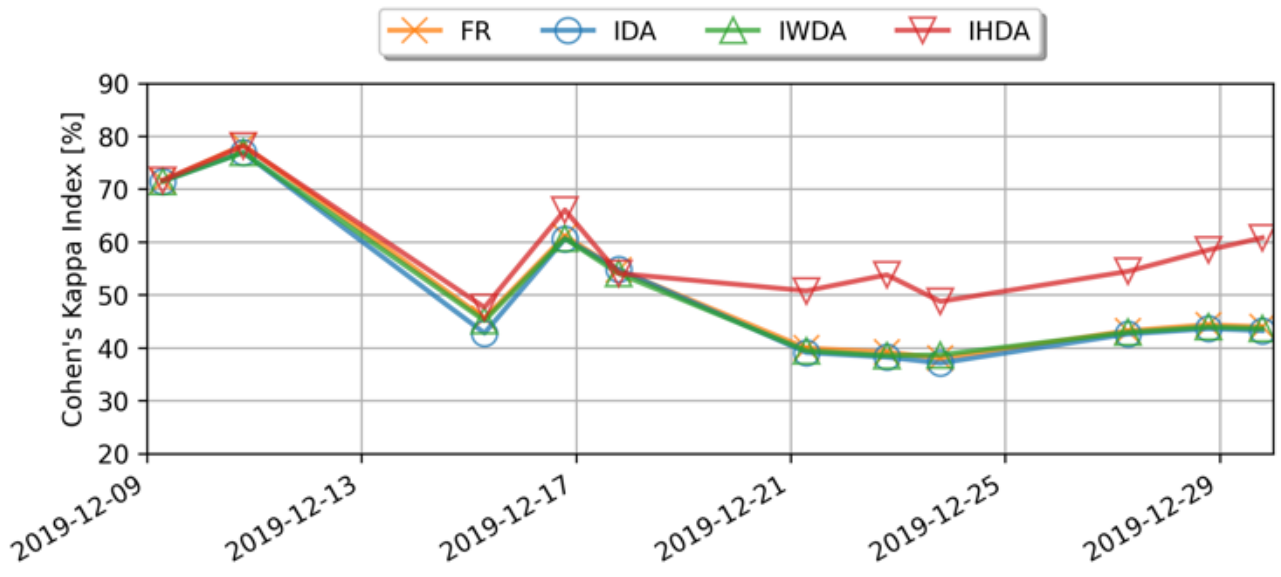

(b) Cohen's κ index

Figure 7. CSI and κ scores.

This emphasizes the complexity of the flood dynamics in the floodplain, and advocates for the further addition of the S2-derived flood observations in the DA. Such a combination of S1 and S2 images is rarely possible due to the unavailability of S2 images during a flood event because of cloud cover problem. Lastly, the 5th row of Figure 6 shows an overall improvement spread out over the five subdomains, as the amount of overprediction regions are significantly reduced. This is due to the state corrections applied at the timesteps between the two flood peaks. Figure 7 depicts the CSI (left column) and the κ index (right column) yielded by FR and DA experiments at all S1 overpass times for the 2019 event with the same color code used previously. These confirm the merits of the DA strategy in IHDA, especially for the representation of the floodplain dynamics and during flood recess.

## V. Conclusions and Perspectives

This study presents the merits of assimilating 2D flood extent observations derived from remote sensing Sentinel-1 SAR images with an Ensemble Kalman Filter implemented on the 2D hydrodynamics model TELEMAC-2D. The flood extent information is expressed in terms of Wet Surface Ratio or WSR computed over defined sensitive subdomains of the floodplain. The WSR is assimilated jointly with in-situ water level observations. The study was carried out over the Garonne Marmandaise catchment, focusing on the flooding event in 2019. Four experiments were realized; one in free run mode and three in DA mode. The control vector gathers friction and forcing correction, and is augmented with correction of the hydraulic state in subdomains of the floodplains (IHDA experiment) that constitute the innovative strategy of this work. All of the DA experiments were implemented by a cycled EnKF with an 18-hour assimilation window sliding with 6-hour overlapping. The simulation results were comprehensively assessed with 1D and 2D metrics with respect to assimilated data as well as with respect to independent flood extent, derived from Sentinel-2 optical imagery data.

The first DA experiment (IDA) involves only in-situ observations whereas the second one (IWDA) assimilates both in-situ observations and WSR observations derived from 2D S1 flood extent maps. These two experiments focus on the sequential correction of friction coefficients and inflow discharge. The spotlight of the article is the IHDA experiment, which not only assimilates both types of observations (similar to IWDA), but also handles a dual state-parameter estimation within the EnKF, by treating inflow discharge and friction coefficients as well as the hydraulic state variable in five particular floodplain subdomains, representing evapotranspiration and/or ground infiltration processes that are unavailable in the TELEMAC-2D model.

We have shown that the assimilation of in-situ data in IDA significantly improves the simulation in the river bed, yet, the dynamics in the floodplain remains incorrect with a significant underestimation of the flood. Indeed, the in-situ observations located in the river bed, do not provide information on the dynamics in the floodplain. The assimilation of WSR data in the floodplain, in IWDA, brings no significant improvement in the river bed and no significant improvement in the flood plain when only upstream forcing and friction coefficients are corrected in the river bed. Indeed, the dynamics of the floodplain is not sensitive to model parameters that are accounted for in its limited control vector (i.e. river bed friction and discharge). In order to allow for the improvement of the dynamics in the flood plain, the assimilation of WSR data must be associated to the augmentation of the control vector with the hydraulic state in the flood plain. More specifically, it was shown that the correction of the augmented control vector in IHDA allows to better represent the flood peak and to efficiently dry out the floodplain during the recess period. From FR to IHDA, the RMSE computed with respect to in-situ data in the river bed is reduced by up to 77-80% at Marmande, whereas the CSI computed with respect to remote-sensing flood extent maps is improved by up to 5.27 percentage points for this flood event. This study confirms the assertion that a densification of the observing network, especially in the floodplain, with remote sensing data and advanced DA strategy, allows to improve the representation of the dynamics of the flow in the floodplains.

This work relies on the implementation of an advanced DA strategy for TELEMAC-2D, especially the development of the observation operator dedicated to WSR, as well as the definition of the associated augmented control vector. Yet, it should be noted that the definition of the subdomains in the floodplain over which the hydraulic state is uniformly corrected, requires a deep understanding of the dynamics of the flood, and is thus not straightforward. This aspect could be further investigated, for instance based on a global sensitivity analysis with respect to the hydraulic state but also to other sources of uncertainty such as topography, especially in the downstream area. Indeed, the same dual state-parameter estimation approach could be applied to correct the bathymetry and topography provided that the size of the uncertainties is reduced, for instance working with a spatially uniform correction or a correction that is only projected onto a limited number of principal components of the errors. In this perspective, we aim to consider using high- and very-high-resolution topography as additional inputs to the model. The use of other imagery datasets (e.g. Landsat-8 and Landsat-9) can also be investigated. In the present work, the combination between remote-sensing data with regards to S1 and S2 data requires further investigation as it seems that the improvements made using S1-derived flood extent maps does not necessarily lead to an improvement with regards to S2-derived flood extents. This could stem from the differences between the S1 and S2 measurement, and the flood extent mapping algorithm. In addition, the identification of S1 or S2 exclusion maps—which signify the reliability of the detected flooded and non-flooded regions—should also be considered taking into account the limitations of each data source.

Lastly, a major perspective of this work stands in the potential non-gaussianity of the WSR observations. This limitation can amount to a loss of optimality of the EnKF which relies on the assumption that the observational error follows a gaussian distribution. On-going work, based on a rich literature based on a change of variable to transform the non-gaussian error into gaussian errors (widely known as Gaussian anamorphosis) is ongoing and yields promising early results.

## ACKNOWLEDGEMENT

This work was supported by CNES, CERFACS and SCO-France. The authors gratefully thank the EDF R&D for providing the Telemac-2D model for the Garonne River, and the SCHAPI, SPC Garonne-Tarn-Lot and SPC Gironde-Adour-Dordogne for providing the in-situ data.

## REFERENCES

[1] Dasgupta, A., Hostache, R., Ramsankaran, R. A. A. J., Grimaldi, S., Matgen, P., Chini, M., ... & Walker, J. P. (2021). Earth Observation and Hydraulic Data Assimilation for Improved Flood Inundation Forecasting. In *Earth Observation for Flood Applications* (pp. 255-294). Elsevier.

[2] Giustarini, L., Matgen, P., Hostache, R., Montanari, M., Plaza, D., Pauwels, V. R. N., ... & Savenije, H. H. G. (2011). Assimilating SAR-derived Water Level Data into a Hydraulic Model: A Case Study. *Hydrology and Earth System Sciences*, 15(7), 2349-2365.

[3] García-Pintado, J., Mason, D. C., Dance, S. L., Cloke, H. L., Neal, J. C., Freer, J., & Bates, P. D. (2015). Satellite-supported Flood Forecasting in River Networks: A Real Case Study. *Journal of Hydrology*, 523, 706-724.

[4] Matgen, P., Montanari, M., Hostache, R., Pfister, L., Hoffmann, L., Plaza, D., ... & Savenije, H. H. G. (2010). Towards the Sequential Assimilation of SAR-Derived Water Stages Into Hydraulic Models using The Particle Filter: Proof Of Concept. *Hydrology and Earth System Sciences*, 14(9), 1773-1785.

[5] Hostache, R., Chini, M., Giustarini, L., Neal, J., Kavetski, D., Wood, M., ... & Matgen, P. (2018). Near-Real-Time Assimilation of SAR-derived Flood Maps for Improving Flood Forecasts. *Water Resources Research*, 54(8), 5516-5535.

[6] Giustarini, L., Hostache, R., Kavetski, D., Chini, M., Corato, G., Schlaffer, S., & Matgen, P. (2016). Probabilistic Flood Mapping using Synthetic Aperture Radar Data. *IEEE Transactions on Geoscience and Remote Sensing*, 54(12), 6958-6969.

[7] Cooper, E. S., Dance, S. L., García-Pintado, J., Nichols, N. K., & Smith, P. J. (2019). Observation Operators for Assimilation of Satellite Observations in Fluvial Inundation Forecasting. *Hydrology and Earth System Sciences*, 23(6), 2541-2559.

[8] Hervouet, J. M. (2007). *Hydrodynamics of Free Surface Flows: Modelling with the Finite Element Method*. John Wiley & Sons.

[9] Gauckler, P. (1867). *Etudes Théoriques et Pratiques sur l'Ecoulement et le Mouvement des Eaux*. Gauthier-Villars.

[10] Besnard, A., & Goutal, N. (2011). Comparison between 1D and 2D Models for Hydraulic Modeling of a Floodplain: Case of Garonne River. *Houille Blanche-Revue Internationale De L'Eau*, (3), 42-47.

[11] Nguyen, T. H., Delmotte, A., Fatras, C., Kettig, P., Piacentini, A., & Ricci, S. (2021). Validation and Improvement of Data Assimilation for Flood Hydrodynamic Modelling Using SAR Imagery Data. In *Proceedings of the 2020 TELEMAC-MASCARET User Conference,* October 2021 (pp. 100-108).

[12] Nguyen, T. H., Ricci, S., Fatras, C., Piacentini, A., Delmotte, A., Lavergne, E., & Kettig, P. (2022). Improvement of Flood Extent Representation with Remote Sensing Data and Data Assimilation. *IEEE Transactions on Geoscience and Remote Sensing*, DOI: 10.1109/TGRS.2022.3147429.

[13] Nguyen, T. H., Ricci, S., Piacentini, A., Fatras, C., Kettig, P., Blanchet, G., Pena Luque, S., & Baillarin, S. (2022). Assimilation of SAR-derived Flood Observations for Improving Fluvial Flood Forecast. In *Proceedings of the 14th International Conference on Hydroinformatics (HIC2022)*, July 2022. arXiv:2205.08471.

[14] Nguyen, T.H., Ricci, S., Piacentini, A., Fatras, C., Kettig, P., Blanchet, G., Pena Luque, S., & Baillarin, S., (2022). Dual State-Parameter Assimilation of SAR-derived Wet Surface Ratio for Improving Fluvial Flood Reanalysis. *Earth and Space Science Open Archive*, doi:10.1002/essoar.10511831.1.